\documentclass{article}
\usepackage[utf8]{inputenc}

\usepackage{showlabels}

\begin{document}
\title{CYLINDRICALLY SYMMETRIC FIELDS IN GENERAL RELATIVITY}

\author {N. O. Santos $^1$ \thanks{e-mail:nilton.santos@obspm.fr}  $\;$ and Anzhong Wang $^2$ \thanks{e-mail: anzhong$\_$wang@baylor.edu}\\
{\small $^1$Sorbonne Universit\'e, UPMC Universit\'e Paris 06, LERMA, Observatoire de Paris/Meudon,}\\
{\small 5 place Jules Janssen, F-92195 Meudon Cedex France.}\\
{\small $^2$GCAP-CASPER, Physics Department, Baylor University, Waco, TX 76798-7316, USA.}}

\date{February 2023}

\maketitle

\begin{abstract}
We present a brief review of exact solutions of cylindrical symmetric fields in General Relativity produced by different perfect fluid sources. These sources are assumed static, stationary, translating and collapsing. Properties of these fields are discussed and some important open questions are called the attention for future research. 
\end{abstract}

\section{Introduction}
Globally cylindrically symmetric solutions of the Einstein field equations, at first sight, may not seem to be physically relevant since they impose infinitely long sources. Nonetheless, under controlled circumstances, they can provide fairly accurate descriptions of different physical phenomena. Important features of cylindrical systems that can help relativistic astrophysics are, for instance, exact models for rotation and the dragging of spacetime, models of extragalactic jets, gravitational radiation, the effect to the cosmological constant describing dark energy, translating fluids that might model beams of light produce by stars, and many others. For references concerning these features see \cite{BSW,Brito}. 

Cylindrically symmetric fields started to be studied in the realm of General Relativity dating back to 1919, when Levi-Civita (LC) \cite{Levi} obtained the vacuum solution of a static cylindrical vacuum spacetime.
Ever since much has been written by researchers trying to grasp its physical and geometrical interpretations. However, this endeavour proved to be difficult and uncertain. Only in 1958 Marder \cite{Marder} established that the LC solution has two independent parameters usually called $\sigma$ and $a$. Understanding the origin, geometry and physics that lies behind these two parameters is a big challenge in understanding the solution. For small values of $\sigma$, noted by LC himself, the corresponding Newtonian field is the external gravitational field produced by an infinitely long homogeneous line mass with $\sigma$ representing the mass per unit length. In this approximation the parameter $a$ is also associated with the constant arbitrary potential that exists in the Newtonian solution.

In 1979 Bonnor \cite{Bonnor} observed that $a$ is also dressed with a relevant global topological meaning, and cannot be removed by coordinate transformations. There is a series of obstacles and apparently contradictory properties of $\sigma$ to allow possible interpretations. Unexpectedly, the parameter $\sigma$ is the most difficult and elusive to be interpreted and there is a long list of articles dedicated to unveil its meaning. In spite of that, the Newtonian limit in cylindrical models agrees well with observations (see \cite{Fujimoto}, \cite{Hockney} and \cite{Song}). 

Research extended LC to cylindrically stationary vacuum spacetimes, obtained independently, by Lanczos \cite{Lanczos} in 1924 and Lewis \cite{Lewis} in 1932. Its physical interpretation is taking too a long time. In its usual form it is presented by two families, one with four real parameters and the other with complex parameters. Only not long ago (in 1998), it was been proved by MacCallum and Santos \cite{MacCallum} that the four parameters can be reduced to three independent ones. However, the general interpretation of these parameters still remain unclear, and the constraints relating them are unknown.    

The vacuum spacetime produced  by a translating source has been obtained by Griffiths and Santos  in 2009 \cite{Griffiths}. Its extension possessing a cosmological constant has not yet considered.

In 1925 Beck was the first to study cylindrical time-dependent vacuum spacetime. The interpretation of these solutions as representing cylindrical gravitational waves was given by Einstein. These same solutions were later reobtained by Einstein and Rosen \cite{Einstein} in 1937. It is still unknown a collapsing source that can be matched to this solution \cite{Prisco}. 

Our aim here is to present a brief review of the exact gravitational fields produced by different cylindrical perfect fluid sources which are static, rotating, under translation and under gravitational collapse. Some important unsolved questions are addressed too.

\section{Einstein's field equations}

The world constructed by Newton to describe his theory of gravitation had much simplicity. Space is always equal to itself where particles move and act upon each other. Furthermore, the gravitational effects propagate with an infinite large velocity. This concept of space sounded not plausible to Einstein and he came with the brilliant idea that space is the gravitational field meaning the Newtonian space itself \cite{Rovelli}. This needed a better formulation of the field equations. Einstein, after a long struggle, found that the Riemann geometry, first proposed by Gauss and generalized to any dimensions by Riemann, could properly describe the curvature of space produced by different matter distributions. By doing this Einstein managed to express the gravitational field equations in an arbitrary coordinate system.

In Riemann's geometry the parallel transport of a vector is proportional to its curvature which is described by a quantity called the Riemann curvature tensor $R_{\alpha\beta\gamma\delta}$. The justification of calling it curvature lies in the fact that it vanishes if and only if the space is flat.

Einstein \cite{Einstein1} derived the field equations in the year 1915 which are given by
\begin{equation}
\label{a1}
R_{\alpha\beta}-\frac{1}{2}Rg_{\alpha\beta}=\kappa T_{\alpha\beta}. 
\end{equation}
They show how the space curvature, represented by the Ricci tensor $R_{\alpha\beta}=R^{\gamma}_{\alpha\gamma\beta}$, and its scalar $R=g^{\alpha\beta}R_{\alpha\beta}$ are brought about by the source of curvature, the matter distribution given by the energy momentum tensor $T_{\alpha\beta}$. In fact not only space curves but time as well and for this reason we call spacetime where Einstein field equations dwell. The coupling constant $\kappa$ in normalized units, the velocity of light $c=1$ and the Newton's constant $G=1$, values $\kappa=8\pi G/c^4=8\pi$. This system of equations, called Einstein's field equations, constitutes a set of 10 partial differential equations with respect to the metric $g_{\alpha\beta}$. In the vacuum case where $T_{\alpha\beta}=0$, these equations reduce to
\begin{equation}
R_{\alpha\beta}=0. \label{a2}
\end{equation}

When spacetime is deprived of sources producing gravitational fields, and thereby producing curvature, the spacetime is flat and given by the metric $\eta_{\alpha\beta}$, called the Minkowski metric,
\begin{equation}
ds^2= \eta_{\alpha\beta}dx^{\alpha}dx^{\beta}=dt^2-dx^2-dy^2-dz^2, \label{a3}    
\end{equation}
with $t$ being the time coordinate and $x$, $y$ and $z$ the Cartesian coordinates. Hence, the Newtonian limit, which one expects to be obtained from Einstein's field equations is produced when the metric is given by
\begin{equation}
g_{\alpha\beta}=\eta_{\alpha\beta}+f_{\alpha\beta}, \label{a4}
\end{equation}
with $f_{\alpha\beta}$ being a small deviation from the Minkowski spacetime.
    
\section{LC vacuum spacetime}

Einstein's field equations (\ref{a1}) are highly nonlinear thus imposing huge difficulties in finding their solutions. Furthermore, after one found their solutions, even more difficulties arise in interpreting them. In spite of great efforts to grasp their physical meanings as is well shown in \cite{Griffiths1}, the majority or even almost the totality of these solutions are still not well understood given their high nonlinearity. We will consider these difficulties in the following by studying cylindrically symmetric gravitational fields.

Schwarzschild \cite{Schwarzschild} in 1916 obtained the first vacuum solution to (\ref{a2}) describing the spherically symmetric vacuum field with the following line element
\begin{eqnarray}
ds^2&=&g_{\alpha\beta}dx^{\alpha}dx^{\beta} \nonumber \\
&=&\left(1-\frac{2M}{r}\right)dt^2-\left(1-\frac{2M}{r}\right)^{-1}dr^2-r^2(d\theta^2+\sin^2\theta d\phi^2), \label{a5}
\end{eqnarray}
where the spherical coordinates are numbered $x^0=t$, $x^1=r$, $x^2=\theta$ and $x^3=\phi$. In the Newtonian approximation (\ref{a5}) becomes
\begin{equation}
g_{00}=1+2U, \label{a6}
\end{equation}
where $U$ is the Newtonian potential and comparing to (\ref{a5}) we have
\begin{equation}
U=-\frac{M}{r}, \label{a7}    
\end{equation}
hence the only parameter stemming from the integration of the field equations for a spherically symmetric vacuum spacetime is the Newtonian mass $M$.

The second solution of Einstein's field equations (\ref{a2}) was obtained a few years later, 1919, after Schwarzschild presented his in 1916. It corresponds to a cylindrical vacuum spacetime, and can be cast in the form
\begin{equation}
ds^2=g_{\alpha\beta}dx^{\alpha}dx^{\beta}=\rho^{4\sigma}dt^2-\rho^{4\sigma(2\sigma-1)}(d\rho^2+dz^2)-
\frac{1}{a}\rho^{2(1-2\sigma)}d\phi^2 \label{1},    
\end{equation}
where the cylindrical coordinates are numbered $x^0=t$, $x^1=\rho$, $x^2=z$ and $x^3=\phi$. In the Newtonian approximation we have corresponding to (\ref{a6}), 
\begin{equation}
g_{00}=e^{2U}, \label{2}
\end{equation}
and compared to (\ref{1}) for this case we have the Newtonian potential
\begin{equation}
U=2\sigma\ln\rho. \label{3}    
\end{equation}
Hence, from this expression we clearly see that for small values of $\sigma$ it is the Newtonian mass per unit length as produced  by an infinitely long homogeneous line mass, as observed by LC himself. Ever since much has been written by researchers trying to unveil its physics and geometrical interpretations. However, the endeavour proved to be difficult and uncertain.

Only in 1958 did Marder \cite{Marder} establish that the solution (\ref{1}) contains two arbitrary independent parameters, $\sigma$ and $a$, differently from the Newtonian fields in which there is only one independent parameter $\sigma$.

We call the attention that for spherical vacuum case, the Schwarzschild metric, in the relativistic and Newtonian theories there appears just one parameter. This fact already suggests some harder difficulties in understanding cylindrically symmetric fields, which is indeed the case.

In the following we review the main properties and physics that lies behind the LC spacetime, which so far has been grasped up to the present time in a large number of papers. We are aware that these results sometimes appear contradictory since some interpretations collide with others. In fact this is one of our main motivations to deepen in Einstein's theory.

\subsection{Nature of the coordinates of the LC spacetime}

The LC metric given by (\ref{1}) can be written in the form \cite{Herrera}
\begin{equation}
ds^2=\rho^{4\sigma}dt^2-\rho^{4\sigma(2\sigma-1)}\left(d\rho^2+\frac{1}{a_m}dm^2\right)-\frac{1}{a_n}\rho^{2(1-2\sigma)}dn^2,
\label{4}
\end{equation}
where $-\infty<t<\infty$ is the time and $0\leq\rho<\infty$ the radial coordinates, and $\sigma$, $a_m$ and $a_n$ are constants. The nature of the coordinates $m$ and $n$, so far unspecified, depends upon the behaviour of the metric coefficients. Either $a_m$ or $a_n$ can be transformed away by a scale transformation depending upon the behaviour of the coordinates $m$ and $n$, thus leaving the metric with only two independent parameters. In orderto find that behaviour we transform the radius $\rho$ into a proper length radial coordinate $r$ by defining
\begin{equation}
\rho^{2\sigma(2\sigma-1)}d\rho=dr, \label{5}
\end{equation}
thus obtaining
\begin{equation}
\rho=R^{1/\Sigma}, \;\; R=\Sigma r, \;\; \Sigma\equiv 4\sigma^2-2\sigma+1, \label{6}    
\end{equation}
and metric (\ref{4}) becomes
\begin{equation}
ds^2=f(r)dt^2-dr^2-h(r)dm^2-l(r)dn^2, \label{7}    
\end{equation}
with
\begin{equation}
f(r)=R^{4\sigma/\Sigma}, \;\; h(r)=\frac{1}{a_m}R^{4\sigma(2\sigma-1)/\Sigma}, \;\; l(r)=\frac{1}{a_n}R^{2(1-2\sigma)/\Sigma}. \label{8}
\end{equation}
Consider $0<\sigma<1/2$, which implies for this range $h(r)$ diverging when $r\rightarrow 0$ and $l(0)=0$. Then we can interpret $m$ as the axial coordinate $-\infty<z<\infty$ with $a_m=1$ by rescaling $z$ and $n$ as the angular coordinate $\phi$ with the topological identification of every $\phi$ with $\phi+2\pi$,  and the metric (\ref{7}) becomes
\begin{equation}
ds^2=R^{4\sigma/\Sigma}dt^2-dr^2-R^{4\sigma(2\sigma-1)/\Sigma}dz^2-\frac{1}{a}R^{2(1-2\sigma)/\Sigma}, \label{9}    
\end{equation}
where $a_n$ is replaced by $a$.

Now consider $1/2<\sigma<\infty$, which implies $h(0)=0$ and $l(r)$ diverging when $r\rightarrow 0$. Then we can interpret $m$ as the angular coordinate $\phi$ with topological identification of every $\phi$ with $\phi+2\pi$, and $n$ as an axial coordinate $-\infty<z<\infty$ with $a_n=1$ by again rescaling $z$, and metric (\ref{7}) becomes
\begin{equation}
ds^2=R^{4\sigma/\Sigma}-dr^2-R^{2(1-2\sigma)/\Sigma}dz^2-\frac{1}{a}R^{4\sigma(2\sigma-1)/\Sigma}d\phi^2, \label{10}    
\end{equation}
where we replaced $a_m$ for $a$.

In both cases, (\ref{9}) and (\ref{10}), where $\sigma>0$ we have $g_{00}\rightarrow 0$ as $r\rightarrow 0$, indicating an attractive singularity. While assuming $\sigma<0$, we obtain $g_{00}\rightarrow\infty$ as $r\rightarrow 0$, indicating a repulsive singularity.

The invariant quantity, under coordinate transformations, built out of the Riemann curvature tensor given by $\mathcal{R}=R_{\alpha\beta\gamma\delta}R^{\alpha\beta\gamma\delta}$, called the Kretschmann scalar, is a good indicator of singularities. Calculating $\mathcal{R}$ for metrics $(\ref{9})$ and $(\ref{10})$ we obtain
\begin{equation}
\mathcal{R}=\frac{64\sigma^2(2\sigma-1)^2}{\Sigma^3r^4}. \label{11}    
\end{equation}
From (\ref{11}) we see that $\mathcal{R}\rightarrow\infty$ as $r\rightarrow 0$, and $\mathcal{R}\rightarrow 0$ for $\sigma=0$, $1/2$ and $\infty$.

Metric (\ref{9}) for $\sigma=0$ becomes
\begin{equation}
ds^2=dt^2-dr^2-dz^2-\frac{1}{a}r^2d\phi^2, \label{12}    
\end{equation}
representing the Minkowski spacetime when $a=1$ in cylindrical polar coordinates $(r,z,\phi)$. However if (\ref{12}) has $a>0$ there is an angle deficit of $2\pi\delta$ given by $\delta=1-1/\sqrt{a}$, producing flat spacetime everywhere except along the axis $r=0$, which is interpreted as a cosmic string. The deficit can represent the tension of the string with mass per unit length $\mu=\delta/4$. If there is an angle excess $a<1$ it would represent a cosmic string under compression with $\mu<0$. Hence the constant $a$ is directly linked to the gravitational analogue of the Aharonov-Bohm effect \cite{Dowker}. This effect shows that gravitation depends on the topological structure of spacetime giving rise to an angular deficit $\delta$ as in the electromagnetic Aharonov-Bohm effect, where a (classical) non-observable quantity (the vector potential) becomes observable (part of it) through a quantum non-local effect. Its gravitational analogue allows a (Newtonian) non-observable quantity (the additional constant to the Newtonian potential) to become observable in the relativistic theory through the angular deficit in strings. For a review in cosmic strings see \cite{Hindmarsh}. 

In the case $\sigma=1/2$ the two metric coefficients $h$ and $l$ in (\ref{8}) are constants, so both $a_m$ and $a_n$ can be set to unity. Then neither $m$ nor $n$ is entitled to be an angular coordinate, and the three coordinates $(r,m,n)$ are better visualized as Cartesian coordinates $(x,y,z)$. Hence metric (\ref{4}) can be written as
\begin{equation}
ds^2=z^2dt^2-dx^2-dy^2-dz^2, \label{13}    
\end{equation}
which is the static plane symmetric vacuum spacetime obtained by Rindler \cite{Silva,Griffiths,Rindler}.

In the next section we calculate the circular geodesics for the above spacetime to try to get further understanding of the properties of different values that $\sigma$ can attain for $\sigma\ge 0$, since $\sigma<0$ would in some way correspond to negative mass densities.

\subsection{Geodesics of LC spacetime}

For the metrics (\ref{9}) and (\ref{10}) the circular geodesics \cite{Silva1} have $\dot r=\dot z=0$ and 
$g_{tt,r}{\dot t}^2+g_{\phi\phi,r}{\dot\phi}^2=0$, where the dot stands for differentiation with respect to the affine parameter $s$. The geodesic angular velocity is defined by $\omega={\dot \phi}/{\dot t}$, and the only non zero component of its velocity is the tangential one, given by $W^{\phi}=\omega\sqrt{g_{tt}}$, with it modulus defined by $W^2=W^{\alpha}W_{\alpha}$ \cite{Herrera1}.

In the case $0\leq\sigma<1/2$, from (\ref{9}) we obtain
\begin{eqnarray}
\omega^2&=&\frac{2\sigma}{1-2\sigma}aR^{2(4\sigma-1)/\Sigma}, \label{14} \\
W^2&=&\frac{2\sigma}{1-2\sigma}. \label{15}
\end{eqnarray}
We note that for a given $\sigma$ the velocity $W$ in (\ref{15}) is the same for all circular geodesics, in agreement with the corresponding Newtonian gravitation. Furthermore, we see that $W$ increases monotonically with $\sigma$, that is from $\sigma=0$ producing $W=0$, to $\sigma=1/4$ attaining $W=1$, the speed of light, and finally $\sigma=1/2$ producing geodesics with $W=\infty$. For small $\sigma$ and $a=1$ from (\ref{14}) and (\ref{15}) we obtain the Newtonian limit $W=\omega r$.

In the case $1/2<\sigma<\infty$ from (\ref{10}) we obtain
\begin{eqnarray}
\omega^2&=&\frac{1}{2\sigma-1}aR^{8\sigma(1-\sigma)/\Sigma}, \label{16}\\
W^2&=&\frac{1}{2\sigma-1}. \label{17}
\end{eqnarray}
With $\sigma$ increasing beyond $1/2$, we note from (\ref{16}) that $W$ diminishes, attaining $W=1$ for $\sigma=1$ and $W=0$ for $\sigma=\infty$.

In other words, the circular geodesics are timelike when either $0<\sigma<1/4$ or $\sigma>1$, are lightlike when $\sigma=1/4$ or $1$, and are spacelike when $1/4<\sigma<1$.

If we redefine $\sigma$ by
\begin{equation}
\sigma=\frac{1}{4\bar\sigma}, \label{18}
\end{equation}
then metric (\ref{10}) with a rescaling of its coordinates becomes (\ref{9}), hence (\ref{15}) and (\ref{16}) reduce to (\ref{13}) and (\ref{14}). This means that the parameter range $1/2<\sigma<\infty$ is equivalent to the range $0<\sigma<1/2$ and the coordinates $z$ and $\phi$ switching their nature.

Hence we might have the following picture for the different values of $\sigma$. For small values of $\sigma$ the metric (\ref{9}) with $t$ and $r$ constants describes cylindrical surfaces with $\phi$ as the  periodic coordinate. As $\sigma$ increases the cylindrical surfaces open out and  become an infinite plane for $\sigma=1/2$. For values of $\sigma$ bigger than $1/2$ the coordinate $z$ becomes periodic forming new cylindrical surfaces perpendicular to previous ones for $0<\sigma<1/2$.

Another interesting geodesic is the one that describes the motion of the particle along the axis of symmetry $z$. These geodesics calculated with (\ref{9}) (we restrict the calculation of this metric since (\ref{10}) is equivalent) produce
\begin{equation}
{\ddot z}=\frac{4\sigma(1-2\sigma){\dot r}{\dot z}}{\Sigma r}. \label{19}
\end{equation}
It means that particles increase their speed along $z$ when distancing radially from the axis, while diminishing their axial speed when moving radially towards the axis. This result indicates that a force parallel to the $z$ axis appears. In the flat case $\sigma=0$ such an effect vanishes, bringing out its non-Newtonian nature. We further discuss this weird geodesic property in the section containing the Lewis spacetime.

For radial geodesics it has been shown \cite{Herrera1} that there exist timelike particles approaching $z$ that are reflected at $r=r_{min}$ and move outwards until attaining $\dot r=0$ at $r=r_max$ repeating endlessly this trajectory. This motion is called confinment of test particles.

In the next section we see some further limits that LC metric satisfies

\subsection{LC spacetime as a limit of the $\gamma$ spacetime}

In cylindrical coordinates, static axially symmetric solutions of Einstein's vacuum equations are given by the Weyl metric \cite{Weyl}
\begin{equation}
ds^2=e^{2\lambda}dt^2-e^{-2\lambda}[e^{2\mu}(d\rho^2+dz^2)+\rho^2d\phi^2], \label{20}    
\end{equation}
with $\lambda(\rho,z)$ and $\mu(\rho,z)$ satisfying
\begin{equation}
\lambda_{,\rho\rho}+\frac{1}{\rho}\lambda_{,\rho}+\lambda_{zz}=0, \label{21}    
\end{equation}
and
\begin{eqnarray}
\mu_{,\rho}&=&\rho(\lambda^2_{,\rho}-\lambda^2_{,zz}), \label{22} \\
\mu_{,z}&=&2\rho\lambda_{,\rho}\lambda_{,z}, \label{23}
\end{eqnarray}
where the comma stands for partial derivation. Synge writes \cite{Synge}, as the most amazing fact, that (\ref{21}) is just the Laplace equation for $\lambda$ in the Euclidean space. Metric (\ref{20}) with the general solution of (\ref{21}), (\ref{22}) and (\ref{23}) is usually referred to as the $\gamma$ metric, and the corresponding functions $\lambda$ and $\mu$ are given by
\begin{eqnarray}
e^{2\lambda}&=&\left(\frac{R_++R_--2m}{R_++R_-+2m}\right)^{\gamma}, \label{24} \\
e^{2\mu}&=&\left[\frac{(R_++R_-+2m)(R_++R_--2m)}{4R_+R_-}\right]^{\gamma^2}, \label{25}
\end{eqnarray}
where
\begin{equation}
R^2_{\pm}=\rho^2+(z\pm m)^2,  \label{26}  
\end{equation}
and $\gamma$ and $m$ are two integration constants. These solutions were first found by Bach and Weyl in 1922 \cite{Bach}. Calculating its Newtonian potential (\ref{3}) we obtain
\begin{equation}
U=\gamma\ln \left(\frac{R_-+z-m}{R_++z+m}\right), \label{27}    
\end{equation}
which corresponds to a potential due to a line segment of length $2l$ and mass per unit length $\gamma/2$ symmetrically distributed along the $z$ axis. Hence the total mass $M$ of the line segment is $M=\gamma m$. The particular case $\gamma= 1$ corresponds to the Schwarzschild metric. This can be seen by taking Erez-Rosen coordinates \cite{Erez} defined by
\begin{equation}
\rho^2=(r^2-2mr)\sin^2\theta, \;\; z=(r-m)\cos\theta, \label{28}     
\end{equation}
and the $\gamma$ metric becomes
\begin{equation}
ds^2=-Fdt^2-\frac{1}{F}[Gdr^2+Hd\theta^2+(r^2-2mr)\sin^2\theta d\phi^2], \label{29}    
\end{equation}
where
\begin{eqnarray}
F&=&\left(1-\frac{2m}{r}\right)^{\gamma}, \label{30} \\
G&=&\left(\frac{r^2-2mr}{r^2-2mr+m^2\sin^2\theta}\right)^{\gamma^2-1}, \label{31} \\
H&=& \frac{(r^2-2mr)^{\gamma^2}}
{(r^2-2mr+m^2\sin^2\theta)^{\gamma^2-1}}. \label{32}
\end{eqnarray}
Now we can easily check that for $\gamma=1$ the metric (\ref{29}) reduces to the Schwarzschild metric.

If we want to compare the $\gamma$ metric to the LC metric, in the limit when its length segment $m\rightarrow\infty$, one notices that by taking this limit in (\ref{24}) and (\ref{25}) the metric diverges. For this reason one can use the Cartan scalars approach to obtain a finite limit. These scalars are the components of the Riemann tensor and its covariant derivatives calculated in a constant frame. Two metrics are equivalent if and only if there exist coordinate and Lorentz transformations which transform the Cartan scalars of one of the metrics into the Cartan scalars of the other. Although the Cartan scalars provide a local characterization of the spacetime, global properties such as topological defects do not appear in them. By doing all this procedure one can prove that locally in the limit $m\rightarrow\infty$ the $\gamma$ metric is the same as the LC metric. Details of these long calculations are given in \cite{Herrera2}. Here we come to an interesting and, so far, weird result showing how apparently unexpected results can stem from long known results like the Schwarzschild and the LC solutions. When the density per unit length of the rod $\sigma=\gamma/2$ has the value $\gamma=1$, or the mass density per unit length $\sigma=1/2$, it becomes the Schwarzschild spherically symmetric spacetime, and in the limit $m\rightarrow\infty$, it becomes the Rindler static plane symmetric vacuum spacetime. This is a remarkable result.

For all the different limiting metrics that LC spacetime can undergo see \cite{Herrera2}. The limits for the circular geodesics of the $\gamma$ spacetime to the LC spacetime are well studied in \cite{Herrera1}.

In the next section we make a brief review of possible sources to the LC spacetime.

\subsection{Sources producing LC spacetime}
The LC spacetime, as we saw, contains two essential constants denoted by $a$ and $\sigma$. The constant $a$ refers to an angle deficit or excess that produces cosmological strings. However, it is $\sigma$ that presents the most serious obtacles to its interpretation. Indeed, for small values $0<\sigma<1/4$, LC describes the spacetime generated by an infinite line mass, with $\sigma$ as mass per unit coordinate length. When $\sigma=0$ the spacetime is flat. However, circular timelike geodesics only exist for $0<\sigma<1/4$, becoming null for $\sigma=1/4$ and being spacelike for $\sigma>1/4$. Furthermore, as the value of $\sigma$ increases from $1/4$ to $1/2$ the corresponding Kretchmann scalar (\ref{11}) diminishes monotonically and vanishes when  $\sigma = 1/2$, implying that the spacetime is flat.

Thus, the question is what does the LC metric represent outside the range $0\leq \sigma\leq 1/4$?

First, we observe that there are known physically satisfactory fluid sources for the LC spacetime satisfying boundary conditions for both ranges of $\sigma$ (see for example \cite{Herrera3}). On the other hand the fact that the scalar Kretschmann decreases with increasing $\sigma$ may not be associated with the strength of the gravitational field. Instead it could be associated with the acceleration of a test particle that would measure more suitably the strength of the gravitational field, which is the case for $1/4<\sigma<1/2$. 

A possible interpretation of the LC solution is a spacetime generated by a cylinder whose radius increases with $\sigma$ and tends to infinity as $\sigma$ approaches $1/2$. This interpretation suggest that when $\sigma=1/2$ the cylinder becomes a plane.

It might be instructive to consider a static cylinder filled with anisotropic perfect fluid and calculate its mass per unit length by using the junction conditions to its exterior LC spacetime. The Whittaker formula \cite{Whittaker} for the active gravitational mass per unit length $\nu$ of the static distribution of perfect fluid with energy density $\mu$ and principal stresses $P_r$, $P_z$ and $P_{\phi}$ inside a cylinder of surface $S$ is
\begin{equation}
\nu=2\pi\int^{r_S}_0(\mu+P_r+P_z+P_{\phi})\sqrt{-g}dr. \label{33}
\end{equation}
Considering a static cylindrically symmetric metric
\begin{equation}
ds^2=Adt^2-dr^2-Cdz^2-Dr^2d\phi^2, \label{34}    
\end{equation}
in which $A$, $C$ and $D$ are only functions of $r$, and from Einstein's field equations we obtain
\begin{equation}
\frac{A_{,rr}}{A}-\frac{1}{2}\frac{A_{,r}}{A}\left(\frac{A_{,r}}{A}-\frac{2}{r}-\frac{C_{,r}}{C}-\frac{D_{,r}}{D}\right)=
\kappa(\mu+P_r+P_z+P_{\phi}). \label{35}
\end{equation}
Substituting (\ref{35}) into (\ref{33}) we have the simple expression
\begin{equation}
\nu=\frac{1}{4}\left(\frac{A_{,r}}{A}\sqrt{-g}\right)_S, \label{36}    
\end{equation}
where regularity on the axis of symmetry \cite{Herrera3} have been assumed. Now taking the LC metric (\ref{9}) for the exterior spacetime of the cylindrical surface $S$, and imposing that it satisfies Darmois\' junction conditions, \cite{Darmois} and \cite{Herrera3}, with its interior spacetime described by (\ref{34}) it amounts to impose the continuity of the metric functions and its derivatives on $S$. By doing so, from (\ref{36}) we obtain
\begin{equation}
\nu=\frac{\sigma}{\sqrt{a}}, \label{37}    
\end{equation}
where $a$ is the constant defined in (\ref{9}). From (\ref{37}) we have that when $a>1$, there is a topological angle deficit, then $\nu<\sigma$, and if $a<1$ there is an angle excess producing $\nu>\sigma$. However, when there is no topological defect, $a=1$, then $\nu$ can in fact be interpreted as mass per unit length of its source. Furthermore, since with cylindrical sources no black holes are formed, one might conclude that the minimum mass per unit length to form a black hole satisfies the constraint $\nu>1/2$. This result would fulfil the present knowledge of black holes that a lower mass limit is required for its formation. We are aware that $\nu$ is model depended and cannot be given a general meaning, but nonetheless it fulfils some of the expected properties.

A last comment on sources for the LC spacetime is the fact that conformally flat static sources do not admit Darmois' matching conditions satisfied for an exterior LC spacetime. This result is proved in \cite{Herrera4}. Conformal flatness of a metric means that its corresponding Weyl tensor vanishes. The interpretation of the Weyl tensor is that it describes the purely gravitational characteristics of the source. This interesting result means that a static cylindrical source deprived of its purely gravitational character cannot be smoothly matched to the exterior LC spacetime. It is conspicuous that for spherical symmetry this result does not stand, since there are conformally flat static spherical sources matched to the Schwarzschild spacetime \cite{Herrera5}.

\section{Lewis vacuum spacetime}
The extension of the LC static cylindrically symmetric vacuum spacetime to a stationary cylindrically symmetric vacuum spacetime was obtained independently by Lanczos in 1924 \cite{Lanczos} and Lewis in 1932 \cite{Lewis}. We consider the spacetime described by the cylindrically symmetric stationary metric
\begin{equation}
ds^2=fdt^2-2kdtd\phi-e^{\mu}(dr^2+dz^2)-ld\phi^2, \label{38}    
\end{equation}
where $f$, $k$, $\mu$ and $l$ are functions only of $r$. The ranges of the coordinates are $-\infty<t<\infty$ for the time coordinate, $0\leq r<\infty$ for the radial coordinate, $-\infty < z<\infty$ for the axial coordinate and $0\leq\phi\leq 2\pi$ for the angular coordinate with the hypersurfaces $\phi=0$ and $\phi=2\pi$ being identified. The coordinates are numbered $x^0=t$, $x^1=r$, $x^2=z$ and $x^3=\phi$. The general vacuum solution $R_{\alpha\beta}=0$ for metric (\ref{38}), in the notation given by \cite{Silva} and \cite{Silva1}, is
\begin{eqnarray}
f&=&ar^{1-n}-a\delta^2r^{1+n}, \label{39} \\
k&=&-(1-ab\delta)\delta r^{1+n}-abr^{1-n}, \label{40} \\
l&=&\frac{(1-ab\delta)^2}{a}r^{1+n}-ab^2r^{1-n}, \label{41} \\
e^{\mu}&=&r^{(n^2-1)/2}, \label{42}
\end{eqnarray}
with
\begin{equation}
\delta=\frac{c}{an}. \label{43}    
\end{equation}
The constants $n$, $a$, $b$ and $c$ can be either real or complex, and the corresponding solutions belong to the Weyl class or Lewis class, respectively. For the Lewis class these constants are given by
\begin{eqnarray}
n&=&im, \label{44} \\
a&=&\frac{1}{2}(a_1+b_1)^2, \label{45} \\
b&=&\frac{a_2+ib_2}{a_1+ib_1}, \label{46} \\
c&=&\frac{m}{2}(a_1^2+b_1^2), \label{47}
\end{eqnarray}
where $m$, $a_1$, $a_2$, $b_1$ and $b_2$ are real constants satisfying
\begin{equation}
a_1b_2-a_2b_1=1. \label{48}    
\end{equation}
The metric coefficients (\ref{39}-\ref{42}) with (\ref{44}-\ref{47}) become
\begin{eqnarray}
f &=& r(a_1^2-b_1^2)\cos(m\ln r)+2ra_1b_1\sin(m\ln r), \label{49}, \\
k&=& -r(a_1a_2-b_1b_2)\cos(m\ln r)-r(a_1b_2+a_2b_1)\sin(m\ln r), \label{50} \\
l&=& -r(a_2^2-b_2^2)\cos(m\ln r)-2ra_2b_2\sin(m\ln r), \label{51} \\
e^{\mu}&=& r^{-(m^2+1)/2}. \label{52}
\end{eqnarray}

A simple deduction of the Lewis metric, where one does not need to consider complex parameters to obtain the Lewis class is given in \cite{Gariel}. There a physical interpretation of the field equations is also provided, which permits to have some understanding of the four parameters appearing in the Lewis solution (\ref{39}-\ref{42}). Another derivation of the Lewis metric is given in \cite{MacCallum}. It is found that three parameters are essential, of which one characterizes the local gravitational field, while the remaining two provide information about the topological identification made to produce cylindrical symmetry.

In Newtonian physics the potential due to cylindrical  matter source, being static or stationary, has the same dependence, that is, it depends only on one parameter, the mass per unit length. For the static vacuum cylindrical field in General Relativity the solution is the LC metric, the one we studied in the previous section, revealing two essential parameters, while for the stationary cylindrical rotating source it has, in its usual form four parameters reducible to three essential parameters \cite{MacCallum}.

In the next section we study the meaning of the parameters appearing in the Lewis metric.

\subsection{The parameters of the Weyl class}
The transformation 
\begin{eqnarray}
d\tau&=&\sqrt{a}(dt+bd\phi), \label{53} \\
d\bar\phi &=& -\frac{1}{n}[cdt-(n-bc)d\phi], \label{54},
\end{eqnarray}
casts the metric (\ref{38}) with (\ref{39}-\ref{42}) in the form
\begin{equation}
ds^2=r^{1-n}dt^2-r^{(n^2-1)/2}(dr^2+dz^2)-\frac{r^{n+1}}{a}d\phi^2. \label{55}    
\end{equation}
This is locally the LC metric. Nevertheless, since $\phi=0$ and $\phi=2\pi$ are identified, $\tau$ defined in (\ref{53}) attains a periodic nature unless $b=0$. On the other hand, the new coordinate $\bar\phi$ ranges from $-\infty$ to $\infty$. A more detailed account of this subject can be found in \cite{Stachel}. In order to globally transform the Weyl class of the Lewis metric into the static LC metric we have to make $b=0$ and $c=0$. Note that in this case, from the transformations (\ref{53}) and (\ref{54}), $\tau$ and $\bar\phi$ become true time and angular coordinates. Hence we can say that $b$ and $c$ are responsible for the non-staticity of this family of solutions of the Lewis metric.

As mention previously, the Cartan scalars provide the local characteristics of a metric. They are obtained through the components of the Riemann tensor and its covariant derivatives calculated in a constant frame. Two metrics are equivalent if and only if there is a coordinate and Lorentz transformations which transform the Cartan scalars of one of the metrics into the Cartan scalars of the other. By performing these calculations for both metrics, the LC metric and the Weyl class of the Lewis metric we find that both are equivalent locally and indistinguishable, which confirms the coordinate analysis made in the beginning of this subsection. Furthermore, we showed that only the parameter $n$ curves spacetime for both static and stationary Weyl class metrics. However, we shall see that the two metrics possess very different topological behaviour.

Details of the calculations of the Cartan scalars are given in \cite{Silva1}.

Considering a cylindrical matter source for the Weyl class metric consisting of a rigidly rotating anisotropic fluid, one of Einstein\'s field equations can be integrated. This integration produces
\begin{equation}
fk_{,r}-kf_{,r}=\xi r, \label{56}    
\end{equation} 
where $\xi$ is an integration constant. Calculating the rotation of the source as given in \cite{Silva1} produces the rotation magnitude given by $\xi/(2fe^\mu/2)$. Now using the matching conditions on the surface of the source cylinder as given by Darmois \cite{Darmois} we find
\begin{equation}
c=-\frac{\xi}{2}. \label{57}
\end{equation}
Note that this constant $c$ is different from the speed of light
used in other places of this chapter. In fact, now it measures the rotation of the cylindrical source, as can be seen from (\ref{57}).

In the Newtonian limit, the velocity term is negligible, then from (\ref{57}) $c\approx 0$, and recalling (\ref{39}), we find that
\begin{equation}
f=e^{2U}. \label{58}    
\end{equation}
Then the Newtonian potential is
\begin{equation}
U=2\sigma\ln r+\frac{\ln a}{2}, \label{59}    
\end{equation}
where $\sigma$ is given by
\begin{equation}
\sigma=\frac{1-n}{4}. \label{60}    
\end{equation}
In Newtonian theory, (\ref{59}) is the gravitational potential of an infinite uniform line mass with mass per unit length $\sigma$. The constant $(\ln a)/2$ represents the constant arbitrary potential that exists in the Newtonian solution. The metric (\ref{39}-\ref{42}) has infinite curvature, according to its Cartan scalars, only at $r=0$ for all $n$ except 
$n = \pm 1$, i.e., $\sigma =0$ and $1/2$. Thus the Weyl class metric has a singularity along the axis $r=0$, then we can say that this spacetime is generated by an infinite uniform line source for $0<\sigma <1/2$. 

Considering the static limit of the Weyl class metric when $n=1$ ($\sigma=0)$ and $b=c=0$, we have from (\ref{39}-\ref{42})
\begin{equation}
ds^2=d\tau^2-dr^2-dz^2-\frac{r^2}{a}d\phi^2, \label{61}
\end{equation}
which is the limit of the LC metric when $\sigma=0$. In the previous section it has been pointed that it generates strings when $a>1$ with mass per unit length $\mu =\delta/4$ and how $a$ is directly linked to the gravitational analog of the Aharanov-Bohm effect \cite{Dowker}.

Considering $c=0$ and $n=1$ ($\sigma=0$) in (\ref{39}-\ref{42}) we have
\begin{equation}
ds^2=d\tau^2+2b\sqrt{a}d\tau d\phi-dr^2-dz^2-\left(\frac{r^2}{a}-b^2a\right)d\phi^2, \label{62}    
\end{equation}
producing a locally flat spacetime. In this cas (\ref{62}) represents the exterior spacetime of a spinning string along the axis of symmetry \cite{Jensen} with the same mass per untit length $\mu=\delta/4$ but with an angular momentum $J$ given by
\begin{equation}
J=-\frac{b\sqrt{a}}{4}, \label{63}    
\end{equation}
for $a>1$.

It has been shown \cite{Jensen} that a quantum scalar particle moving around a spinning cosmic string as given by (\ref{62}), exhibits a phase factor proportional to $J$ in its angular momentum. It is a reminiscence of the Aharonov-Bohm effect. It is also worth mentioning that even if $b=0$, an Aharonov-Bohm like effect exists, thought of a different kind, appears (as commented in the static case), since the angular momentum spectrum differs from the usual one, if only $a>1$.

\subsection{Geodesics of Weyl class spacetime}
The Weyl class of solutions is given by (\ref{38}) with (\ref{39}-\ref{42}) and all parameters being real, as shown above. Then for circular geodesics \cite{Herrera6} $\dot r=\dot z=0$, we find that
$f_{,r}{\dot t}^2-2k_{,r}{\dot t}{\dot \phi}-l_{,r}{\dot \phi}^2=0$, where the dot stands for differentiation with respect to an affine parameter $s$. The geodesic angular velocity is defined by $\omega={\dot\phi}/{\dot t}$, and the velocity of the test particle has only two non zero components, $W^t=k\omega/(f^{3/2}-{\sqrt f}k\omega)$ and $W^{\phi}={\sqrt f}\omega/(f-k\omega)$, with 
\begin{equation}
\omega=\frac{c\pm n\omega_0}{n-b(c\pm n\omega_0)}, \;\; W=\frac{\delta r^n\pm W_0}{1\pm\delta r^n W_0}, \label{64} 
\end{equation}
where $\omega_0$ is the LC angular velocity and $W_0$ is the LC tangential velocity,
\begin{equation}
\omega_0^2=\left(\frac{1-n}{1+n}\right)\frac{a^2}{r^{2n}}, \;\; W_0^2=\frac{1-n}{1+n}. \label{65}    
\end{equation}
We note that $\omega$ and $W$ vanish for $\omega_0=\mp c/n$ and $W_0=\mp\delta r^n$, respectively, which are equivalent to say that the free particle in the present case is simply static. This could come about if the ``centrifugal repulsion" balances the gravitational attraction.

The geodesic motion of a particle along the axis of symmetry $z$ for metric (\ref{38}) produces
\begin{equation}
\ddot z=\frac{(1-n^2){\dot r}{\dot z}}{2r}. \label{66}    
\end{equation}
It is interesting to note that for this geodesic the parameters $b$ and $c$, due to the stationarity of spacetime, do not appear in (\ref{66}) and in fact it is indistinguishable from the static limit, the LC spacetime. There is a force that tends to damp the motion along the axis, $\ddot z <0$, whenever the particle approaches the axis, $\dot r<0$, and reverses this tendency in the opposite case. In the flat case, $n=1$ or $\sigma=0$, such an effect vanishes, exposing its non-Newtonian nature.

For quasi-spherical objects it has been shown \cite{Herrera7} that positive radial acceleration can be produced along its axis of symmetry.

It is also worth noticing that the non-Newtonian forces parallel to the $z$ axis, also appear in the field of axially symmetric rotating bodies \cite{Bonnor1}. However the force parallel to $z$ found in \cite{Bonnor1}, unlike the current case, is directly related to the spin of the source. For the Kerr black hole, it is shown that particles produced by the Penrose process, can be ejected from the ergosphere surface, covering the the black hole through repulsive gravitational fields. In this case too, unlike in the Lewis spacetime, the gravitational repulsion is created by the spin of the Kerr black hole. These ejected particles are highly collimated and might be a mechanism for the observed extragalactic jets \cite{Gariel1,Gariel2,Pacheco}.

\subsection{The parameters of the Lewis class}
Using the transformation
\begin{equation}
 d\phi=d{\bar\phi}+\omega dt, \;\; \omega=-\frac{k}{l}, \label{67}   
\end{equation}
the metric (\ref{49}-\ref{52}) can be diagonalized. In order to have an intergal coordinate transformation $\omega$ must be a constant, therefore,
from (\ref{49}-\ref{52}), $m=0$. This implies, from (\ref{44}-\ref{47}) that $n=0$ and $c=0$. Thus the line element becomes
\begin{equation}
ds^2=-\frac{rdt^2}{a_2^2-b_2^2}-\frac{(dr^2+dz^2)}{\sqrt r}-(a_2^2-b_2^2)rd\bar{\phi}^2. \label{68}    
\end{equation}
This is a particular case of the static LC metric with the energy density per unit length $\sigma$ given by (\ref{60}) being equal to $1/4$. Nevertheless, the transformation (\ref{67}) is not global, since the new coordinate $\bar\phi$ ranges from $-\infty$ to $\infty$, instead of ranging from $0$ to $2\pi$ \cite{Stachel}.

Considering, as in the case of the Weyl class, a rigidly rotating anisotropic fluid, one of the Einstein's field equations can be integrated producing (\ref{56}). With (\ref{49}-\ref{52}) the matching conditions given by Darmois \cite{Darmois} yield
\begin{equation}
\xi=-m(a_1^2+b_1^2), \label{69}
\end{equation}
and so
\begin{equation}
c=-\frac{\xi}{2}. \label{70}    
\end{equation}
Hence, in order to have the rotation equal to zero, i.e. $\xi=0$, we need $m=0$ since $a_1^2+b_1^2\neq 0$.

Observe the difference, at this point, between the Weyl class and the Lewis class. In the latter the vanishing of the rotation yields a locally LC spacetime, whereas in the former the vanishing of the rotation does not necessarily imply that the metric can be reduced either globally or locally to a static spacetime.

For the Cartan scalars that produce the local characteristics of a metric, we have the following results for the Lewis class. In the Lewis class as in the Weyl class, only the constant $n$ appears in the Cartan scalars. Nevertheless, now $n$ must be substituted by its complex value $im$ (\ref{44}). However, contrary to the Weyl class, the Cartan scalar for the Lewis class are distinguishable from the LC metric, except for $m=0$. Furthermore, there is no value of $m$ for which the Cartan scalars are all zero, implying at once that the Lewis class does not include the Minkowski spcetime as a special case. This fact implies too that there must be a lower limit to the energy per unit length of its source. The Cartan scalars impose a upper bound on the parameter $m$, given by
\begin{equation}
m\leq\sqrt 3, \label{71}    
\end{equation}
since for larger values of $m$ than this, the singularity is ar $r=\infty$ and not at $r=0$.

\subsection{Sources producing Lewis spacetime}
In a fine paper, well ahead of his time, as observed by Bill Bonnor \cite{Bonnor2}, in 1937 van Stockum \cite{Stockum} completely solved the problem of a rigidly rotating infinitely long cylinder of dust, including the application of adequate boundary conditions. The solution is a remarkable one. The metric for the interior is simple and unique depending on one parameter $w$ in our notation. But, for the vacuum exterior, $r>R$, where $R$ is the coordinate radius of the cylinder, there are three cases depending on the mass per unit length of the interior. For the metric (\ref{38}) we have the following results.

Case $wR<1/2$:
\begin{eqnarray}
f&=&-r\left[2\beta\cosh(2N\ln r)+\frac{\alpha^2+\beta^2}{\alpha}\sinh(2N\ln r)\right], \nonumber \\
k&=&-r\left[\cosh(2N\ln r)+\frac{\beta}{\alpha}\sinh(2N\ln r)\right], \nonumber \\
l&=&\frac{r}{\alpha}\sinh(2N\ln r), \nonumber \\
e^{\mu}&=&\lambda\left(\frac{r}{R}\right)^{(4N^2-1)/2}, \label{72}
\end{eqnarray}
with
\begin{eqnarray}
N&=&\frac{1}{2}(1-4w^2R^2)^{1/2}, \;\; \alpha=\frac{(1-4w^2R^2)^{1/2}}{2w^3R^4}, \nonumber \\
\beta&=&\frac{2w^2R^2-1}{2w^3R^4}, \;\; \lambda=e^{-w^2R^2}. \label{73}
\end{eqnarray}

Case $wR>1/2$:
\begin{eqnarray}
f&=&r\left[2\beta\sin(2N\ln r)+\frac{\alpha^2-\beta^2}{\alpha}\cos(2N\ln r)\right], \nonumber \\
k&=&r\left[\sin(2N\ln r)-\frac{\beta}{\alpha}\cos(2N\ln r)\right], \nonumber, \\
l&=&\frac{r}{\alpha}\cos(2N\ln r), \nonumber \\
e^{\mu}&=&\lambda\left(\frac{r}{R}\right)^{(1-4N^2)/2}, \label{74}
\end{eqnarray}
with
\begin{eqnarray}
N&=&\frac{1}{2}(4w^2R^2-1)^{1/2}, \;\; \alpha=\frac{(4w^2R^2-1)^{1/2}}{2w^3R^4}, \nonumber \\
\beta&=&\frac{2w^2R^2-1}{2w^3R^4}, \;\; \lambda=e^{-w^2R^2}. \label{75}
\end{eqnarray}

For the case $wR=1/2$ one obtains the relations from the limits either from 
$wR<1/2$ or $wR>1/2$ which are equal. The solution for $wR<1/2$ belongs to the Weyl class and its real parameters $n$, $a$, $b$ and $c$ assume the values
\begin{eqnarray}
n&=&(1-w^2R^2)^{1/2}, \;\; a=\frac{(\alpha-\beta)^2}{2\alpha}, \nonumber \\
b&=&\pm\frac{1}{\alpha-\beta}, \;\; c=\frac{\alpha^2-\beta^2}{\alpha}N. \label{76}
\end{eqnarray}

The solution $wR>1/2$ belong to the Lewis class and its real parameters $m$, $a_1$, $a_2$, $b_1$ and $b_2$ assume the values
\begin{eqnarray}
m&=&(4w^2R^2-1)^{1/2}, \;\; a_1=\frac{\beta}{b_1}, \;\; a_2=-\frac{1}{b_1},   \nonumber \\
b_1&=&(-\alpha)^{1/2}, \;\; b_2=0. \label{77}
\end{eqnarray}

For the Weyl class we have the Newtonian mass per unit length given by $\sigma=(1-n)/4$ which implies, for Case $wR<1/2$, from (\ref{76}),
\begin{equation}
\sigma=\frac{1}{4}\left[1-(1-4w^2R^2)^{1/2}\right]. \label{78}    
\end{equation}
Hence (\ref{78}) establishes a lower limit for $\sigma$ in the Lewis class and being $\sigma=1/4$. This value is the frontier between the Weyl class metric and the Lewis class metric, at least for the rotating dust solution obtained by van Stockum \cite{Stockum}.

For the Lewis class metric the Cartan scalars, as it was remarked, do not admit the Minkovski spacetime. This is in accordance with the existence of a lower limit for $\sigma$ in the van Sockum solution \cite{Stockum} $wR>1/2$, since with a lower limit the source cannot be made a vacuum, and therefore the exterior solution cannot be Minkovski.

The Cartan scalars also impose an upper bound on the parameter $m$, given by
\begin{equation}
m\leq\sqrt 3, \label{79}    
\end{equation}
since for values of $m$ larger than this, the singularity is at $r=\infty$ and not in $r=0$. When we substitute this value in (\ref{77}), considering the equality, we have $wR=1$.

Van Stockum solution is studied at length in \cite{Bonnor2}, its properties concerning gravitoelectric and gravitomagnetic fields in \cite{Bonnor1} and \cite{Costa}, its confinement properties in \cite{Opher}, its extension to non-rigid rotation in \cite{Bonnor3}. A range of stationary cylindrically symmetric perfect fluid sources are presented in \cite{Celerier} and an anisotropic cylindrical stationary source can be found in \cite{Debbasch}.

\section{Translation of cylinders}
It has been proved \cite{Griffiths} that the vacuum field produced by a rotating mass cylinder is mathematically closely related to the field produced by a translating mass cylinder along its axis of symmetry. Nonetheless, its physical and geometrical properties differ significantly since the relativistic frame dragging for rotation and translation physically differ considerably.

We assume the general cylindrically symmetric metric with its source translating parallel to its axis of symmetry given by
\begin{equation}
ds^2=Adt^2-2Kdtdz-Bd\rho^2-Cdz^2-D\rho^2d\phi^2, \label{80}    
\end{equation}
with the usual properties of its coordinates and $A$, $K$, $B$, $C$ and $D$ functions only of $\rho$. Making $\rho=e^r$ and after rescaling, the metric (\ref{80}) can be written as
\begin{equation}
ds^2=fdt^2-2kdtdz-e^{\mu}(dr^2+d\phi^2)-dz^2. \label{81}    
\end{equation}
The general vacuum solution for (\ref{81}) 
is the stationary Lewis metric. So, the vacuum solution corresponding to (\ref{80}) is simply the lewis solution with the coordinates $z$ and $\phi$ interchaged. Hence the metric coefficients in (\ref{81}) are the same as in (\ref{39}-\ref{42}).

In spite of the mathematical similarity between the vacuum solutions for the fields produced by rotating and translating cylinder filled with perfect fluid they differ substantially. Unlike the rotating case, the translating cylinder cannot be filled with pressure free dust, as there is nothing here equivalent to a centrifugal force that would prevent the matter collapsing to the axis. The pressure must therefore be nonzero. Furthermore, unlike the rotating case where matter can be rigidly rotating, which means with shear free rotation, the translating matter case if translating rigidly can always be transformed to a frame where the system is static \cite{Griffiths}.

The field of a cylinder of matter that is in translational motion along its length has not been studied in detail, and any differences with the static case are unknown. It is therefore of interest to determine whether or not frames are dragged by motion along the cylinder in a way similar to that in which they are dragged around it.

\section{Gravitational collapse of cylinders}
Gravitational collapse and the emission of gravitational waves has been one of the most important problems in Einstein's theory. However, due to the complexity of Einstein's field equations, even in simple cases it is not well understood. It is well known, due to the Birkhoff's theorem it is not possible to have gravitational radiation in spherically symmetric spacetimes. The next simplest symmetry assumption is cylindrical symmetry. In this sense gravitational collapse of a cylindrical distribution of perfect fluid mass matched to an exterior time depended vacuum field might be important since it can be well stated and might give us hope to obtain exact solutions satisfying matching conditions with parameters that can help us to understand the mechanism of the eventual production of gravitational waves.

For the exterior of a cylindrically anisotropic perfect fluid under gravitational collapse one considers the Beck-Einstein-Rosen metric, \cite{Beck} and \cite{Einstein}, for vacuum (\ref{a2}) given by
\begin{equation}
ds^2=e^{2(\gamma-\psi)}(dT^2-dR^2)-e^{2\psi}dZ^2-e^{-2\psi}R^2d\phi^2, \label{82}
\end{equation}
where $\gamma$ and $\psi$ are functions of $T$ and $R$, and vacuum field equations become
\begin{equation}
\psi_{,TT}-\psi_{,RR}-\frac{\psi_{,R}}{R}=0, \label{83}    
\end{equation}
and
\begin{equation}
\gamma_{,T}=2R\psi_{,T}\psi_{,R}, \;\; \gamma_{,R}=R(\psi_{,T}^2+\psi_{,R}^2). \label{84}    
\end{equation}
Equation (\ref{83}) is the cylindrically symmetric wave equation in an Eucledean spacetime, suggesting the presence of a gravitational wave field. All the set of equations that (\ref{82}) has to satisfy in order to be matched, following Darmois' conditions, to an anisotropic perfect fluid under gravitational collapse are given in \cite{Prisco}. There it is demonstrated that for a shearfree cylindrically symmetric anisotropic fluids if the exterior spacetime is static, i.e. the LC spacetime, the cylindrical source must be static too. However, if the fluid is isotropic and shearfree, only a Robertson-Walker dust interior. It is concluded as well that there is no energy transport in its exterior.

It is still an open question the more general behaviour of a cylindrically collapsing anisotropic fluid producing shear.

Further studies with collapsing dust have been undertaken by considering nonzero shear but with zero expansion \cite{Brito1} and other simplifying assumptions see references in \cite{Brito} and \cite{Brito1}.

\section{Conclusions}

With the beginning of the era of the gravitational wave astronomy, the strong gravitational field regime will be soon explored observationally in various aspects. Theoretical studies can be carried out analytically and/or numerically. In the former, due to the complexity of Einstein's field equations symmetries of spacetimes are often imposed such as spherical, plane and cylindrical \cite{Stephani}. Although they are all ideal models, and in realistic situations any of these may not exist, they do provide solvable problems from which some fundamental issues of physics can be addressed. A good example is the Schwarzschild solution which plays a central role in the studies of black hole physics.

In this brief review concerning different cylindrically symmetrical perfect fluid sources, static, stationary, translating and collapsing, we present what is so far achieved in understanding the physics and geometry. We hope that this brief adventure into cylindrical systems in General Relativity helps to justify its importance and motivate further studies on unsolved issues like the ones that we called the attention in this review.

\end{document}